\documentclass{appolb}

\usepackage[english]{babel}
\usepackage{graphicx}
\usepackage{mathrsfs,amssymb,amsmath,amsopn,bm}
\usepackage{color}
\usepackage{slashed}

\newcommand{\MeV}{\ensuremath{\mathrm{MeV}}}

\newcommand{\bvec}[1]{\ensuremath{\boldsymbol{#1}}}
\newcommand{\rr}{\bvec{r}}
\newcommand{\pp}{\bvec{p}}
\newcommand{\ii}{\ensuremath{\mathrm{i}}}
\newcommand{\dd}{\ensuremath{\mathrm{d}}}

\begin{document}
\title{Dynamics of the chiral phase transition%
\thanks{XXXI Max Born Symposium and HIC for FAIR Workshop}%
}
\author{H.\ van Hees$^{1,2}$, C.\ Wesp$^1$, A.\ Meistrenko$^1$ and C.\ Greiner$^1$
\address{$^1$Institut f{\"u}r Theoretische Physik, Goethe-Universit{\"a}t Frankfurt,
Max-von-Laue-Stra{\ss}e 1, D-60438 Frankfurt, Germany\\ $^2$ Frankfurt
Institute for Advanced Studies, Ruth-Moufang-Stra{\ss}e 1, D-60438 Frankfurt, Germany}
}
\maketitle
\begin{abstract}
  The intention of this study is the search for signatures of the chiral
  phase transition in heavy-ion collisions. To investigate the impact of
  fluctuations, e.g., of the baryon number, at the transition or at a
  critical point, the linear sigma model is treated in a dynamical
  (3+1)-dimensional numerical simulation. Chiral fields are approximated
  as classical mean fields, and quarks are described as quasi particles
  in a Vlasov equation. Additional dynamics is implemented by
  quark-quark and quark-sigma-field interactions. For a consistent
  description of field-particle interactions, a new
  Monte-Carlo-Langevin-like formalism has been developed and is
  discussed.
\end{abstract}

\PACS{PACS numbers come here}
  
\section{Introduction}

In this proceeding we will present our work on the linear $\sigma$-model
off thermal equilibrium.  Section \ref{LinearSigmaModel} describes the
physical model with the numerical implementation described in section
\ref{NumericalImplementation}. Consistency tests and results are
discussed in section \ref{TestCasesAndResults} in which we also discuss
difficulties implicated by only including elastic and mean-field
interactions in such a model. In section
\ref{sectionParticleWaveInteraction} we will relate to this class of
problems and discuss our new method to allow hard and scattering-like
interactions between particles and classical fields.

\section{Linear $\sigma$-model}
\label{LinearSigmaModel}

\begin{figure}[]
  \centering
  \includegraphics[width=0.75\textwidth]{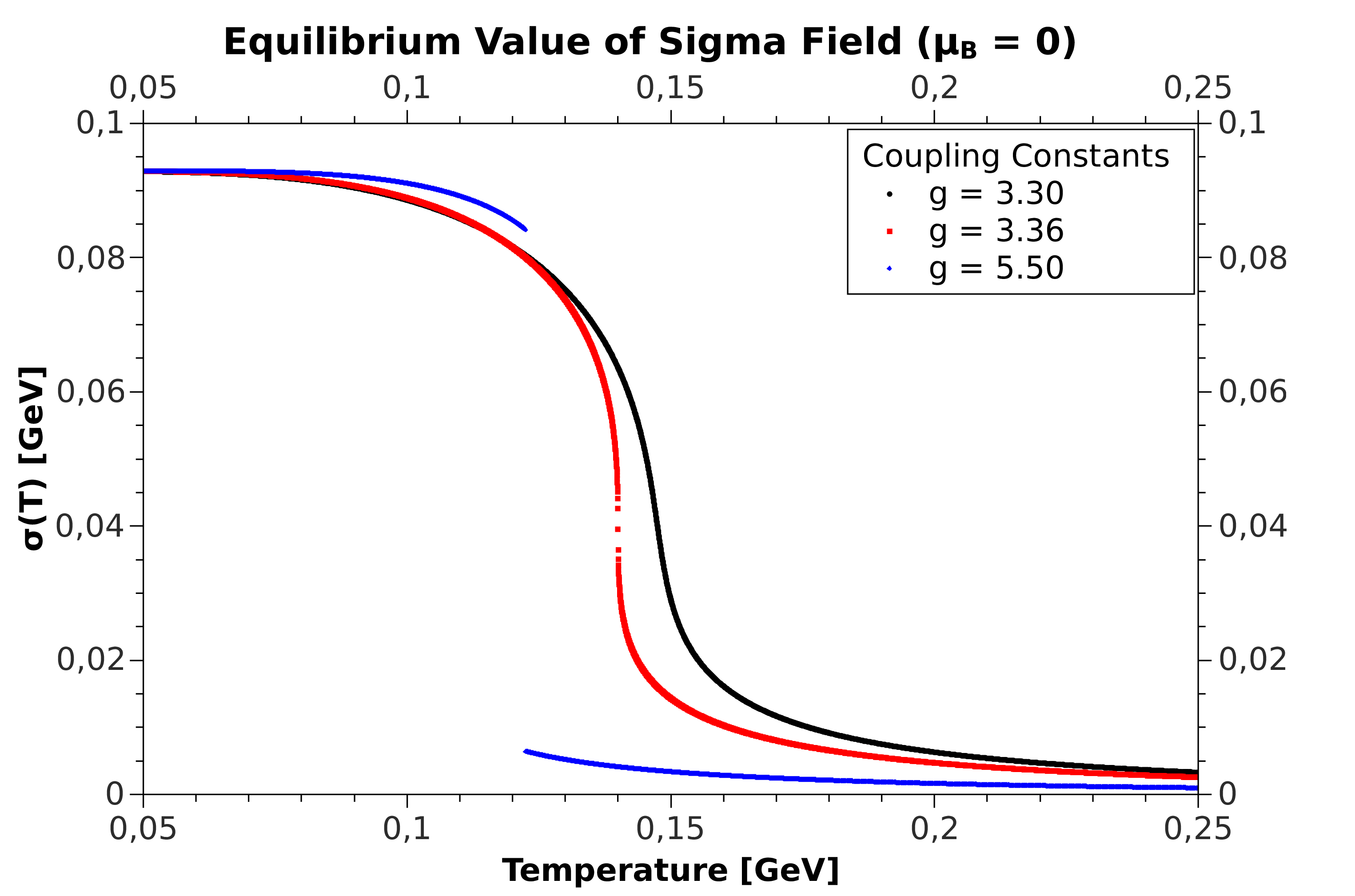}
  \caption{Thermal behaviour of the model when initialised in
    equilibrium. Depending on the coupling strength, the order of the
    phase transition changes.}
 \label{equilibrium1}
\end{figure}
For our numerical studies we use the effective O(4)-linear sigma model
\cite{gell1961nuovo, scavenius2001chiral}, which is motivated by the
chiral symmetry of QCD. It is a model with constituent quarks and
anti-quarks, coupled to the chiral fields $\phi \in \mathbb{R}^4$.  By
spontaneous and explicit symmetry breaking, the chiral fields are
decomposed into the pionic fields $\vec \pi$ and a massive field
$\sigma$, acting as an order parameter. The Lagrangian reads
\begin{equation} 
  \mathscr{L} = \bar \psi \left [ \ii \slashed \partial \ {- g \left(
        \sigma + \ii \vec \pi \cdot \vec \tau \gamma_5 \right) } \right
  ] \psi - \frac{1}{2} \left (\partial_\mu \sigma \partial^\mu \sigma
    + \partial_\mu \vec \pi \partial^\mu \vec \pi \right ) 
  - U\left( \sigma, \vec \pi \right) 
\end{equation} 
with the chiral potential
\begin{equation}
  U \left( \sigma, \vec \pi \right) = \frac{\lambda^2}{4} \left
    (\sigma^2 + \vec \pi^2 - \nu^2 \right)^2 + {U_0(f_\pi, m_\pi,
    \sigma)}. 
\end{equation}
Quarks are dynamically coupled to the meson mean fields via their
dispersion relation:
\begin{alignat}{2} 
  E_q^2 &= p^2 + m_q^2(\bvec x, t), \\
  m_q^2(\bvec x, t) &= g^2 \left [\sigma^2(\bvec x, t) + \pi^2(\bvec x, t) \right],
\end{alignat}
leading to a position and time dependent mass term for the quarks. The
model parameters are chosen to fit the pion mass, $m_\pi=138 \; \MeV$,
and the vacuum expectation value of the order parameter $\langle \sigma
\rangle_0 = f_{\pi} = 93 \; \MeV$ in the vacuum. With an interaction
coupling of $g = 3$-$6$, the constituent-quark mass can be adjusted to
$m_q \approx 330 \; \MeV$.  Depending on $U_0$, chiral symmetry can be
explicitly broken leading to a non-vanishing pion mass, $m_{\pi}$. These
adjustments lead to the parameters $\lambda^2=20$, $g=3$-$6$, $U_0=m^4_\pi
/ \left(4 \lambda^2 \right) - f_\pi^2 m_\pi^2$, and
$\nu^2=f_{\pi}^2-m_{\pi}^2/\lambda^2$. Figure \ref{equilibrium1} shows the thermal
behaviour of the model when it is initialised in a state of kinetic and
chemical equilibrium.

\section{Numerical implementation}
\label{NumericalImplementation}

For the numerical implementation we treat the meson fields on the
mean-field level and the quarks as particles in a Vlasov equation for
their phase-space distribution function $f(t,\bvec{r},\bvec{p})$,
augmented by dissipation and particle production,
 \begin{equation}
\begin{split}
\label{5}
    \left[ \partial_t + \frac{\pp}{E(t, \rr, \pp)} \cdot \nabla_{\rr} -
      \nabla_{\rr} E(t,\rr, \pp) \cdot \nabla_{\pp} \right ] \ f(t,\rr,\pp) \\
    + \text{`Dissipation'} + \text{Particle production}  = 0.
\end{split}
 \end{equation}
The phase-space distribution function for the quarks is treated in a
Monte-Carlo test-particle description,
  \begin{equation}
   f(t, \rr, \pp) = \frac{1}{N_{\text{test}}} \sum_i \delta^{(3)} \left[ \rr
     - \rr_i(t) \right] \ \delta^{(3)} \left [ \pp - \pp_i(t) \right ].
  \end{equation}
The mean-field equations for the meson fields,
\begin{equation}
\begin{split}
    \partial_\mu \partial^\mu \ \sigma + \lambda^2 \left( \sigma^2 +
      \vec \pi^2 - \nu^2 \right) \sigma + g \langle \bar \psi \psi
    \rangle - f_\pi m_\pi^2 \\ 
     + \text{`Dissipation'} + \text{Particle prod.} = 0
\end{split}
\end{equation}
are solved on a 3D grid using the leap-frog algorithm.
\begin{figure}
  \centering
  \includegraphics[width=0.75\textwidth]{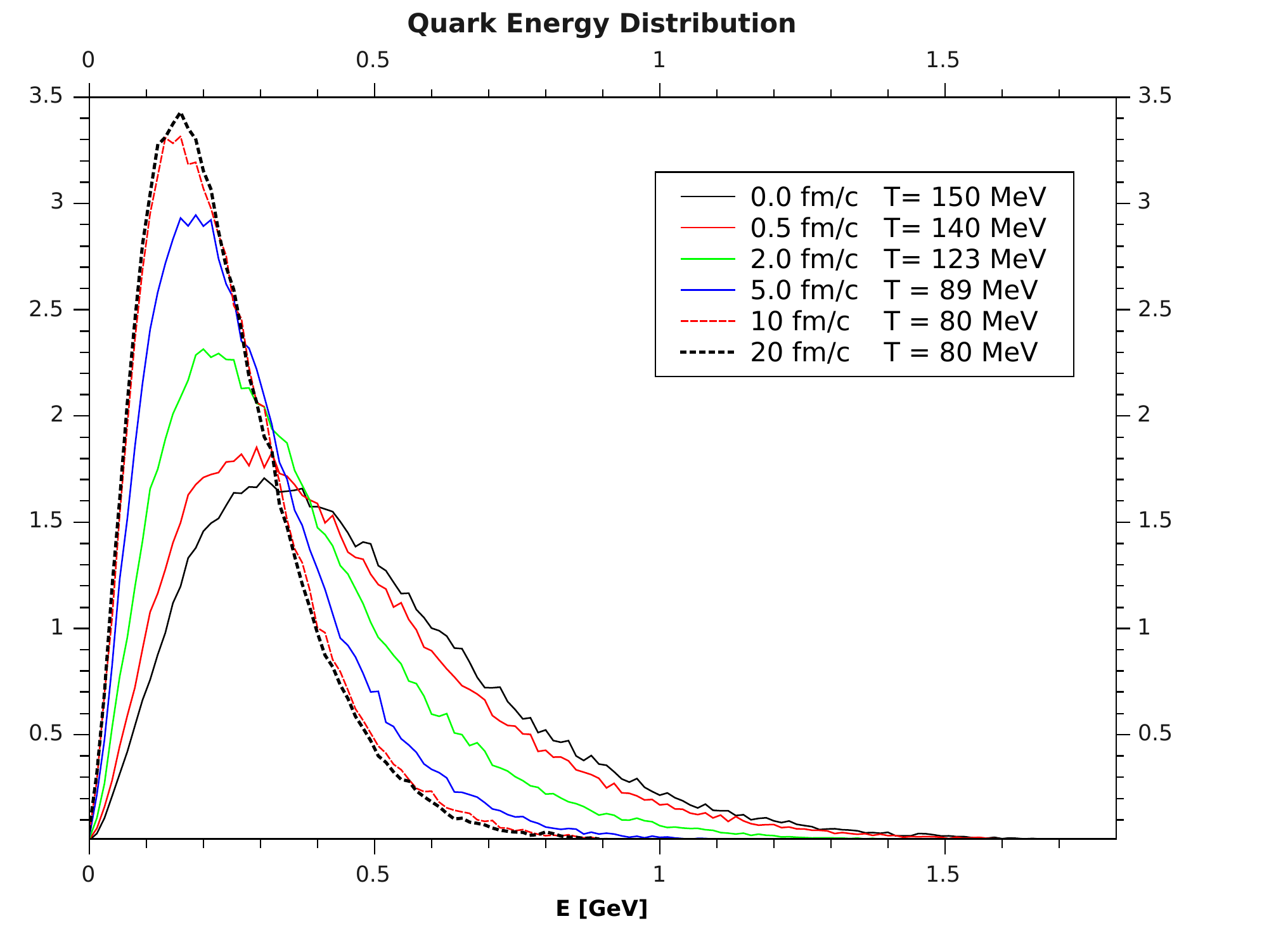}
  \caption{Thermal cooling of the quarks via interaction with a
    heat-bath.}
 \label{cooldown1}
\end{figure}

\subsection{Quark Interactions} 
\label{QuarkInteraction}

So far, the quarks are coupled to the chiral mean fields only via the
Vlasov term. To add additional dynamics and interactions, we implement a
collision term for the quasi-particles. This method is equivalent to the
approach used in the BAMPS model \cite{Xu:2004mz}.

The quark's position space is divided into grid cells $\Delta^3 x$,
allowing all particles within a cell to interact. The interaction
probability is given by
\begin{equation}
 P_{22} = \frac{s}{E_1 E_2} \frac{\sigma_{22}}{N_\textrm{test}} \frac{\Delta t}{\Delta^3 x} \ ,
\end{equation}
with $\sigma_{22}$ denoting an isotropic and constant cross section.

Additionally we want to be able to cool down or to heat up the system.
Implementing a virtual heat bath, the quasi-particles can interact with
thermal particles sampled from a heat bath, allowing the particles to
statistically gain or lose energy. Figure \ref{cooldown1} shows the change of the
system's temperature due to kinetic interactions of the quarks with a
virtual heat bath. In this scenario, the quarks are cooled from $T=150
\; \MeV$ to $T=80 \; \MeV$.
\begin{figure}[]
  \centering
  \includegraphics[width=0.8\textwidth]{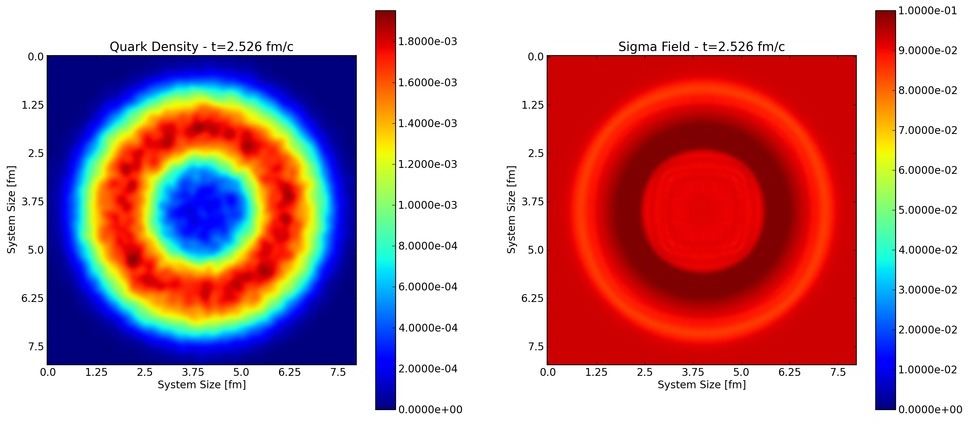}
  \caption{Expansion of a droplet initial condition. The particle
    densities and momenta are sampled according to a Woods-Saxon
    potential.}
 \label{thermalBlobb}
\end{figure}

\section{Test Cases and Results}
\label{TestCasesAndResults}

Several test cases and scenarios have been simulated. In basic
scenarios, we have checked for numerical stability, energy conservation,
vacuum properties and have compared the temperature dependent
equilibrium properties with values from the literature
\cite{nahrgang2011nonequilibrium}.  Figure \ref{thermalBlobb} shows a thermal-droplets
scenario. The system is initialised with a thermal Woods-Saxon potential
to resemble a hot and dense fireball.  Then the system cools down by
expansion and shows shell-like structures of the quark distribution and
the chiral fields.

\begin{figure}[]
  \centering
  \includegraphics[width=0.75\textwidth]{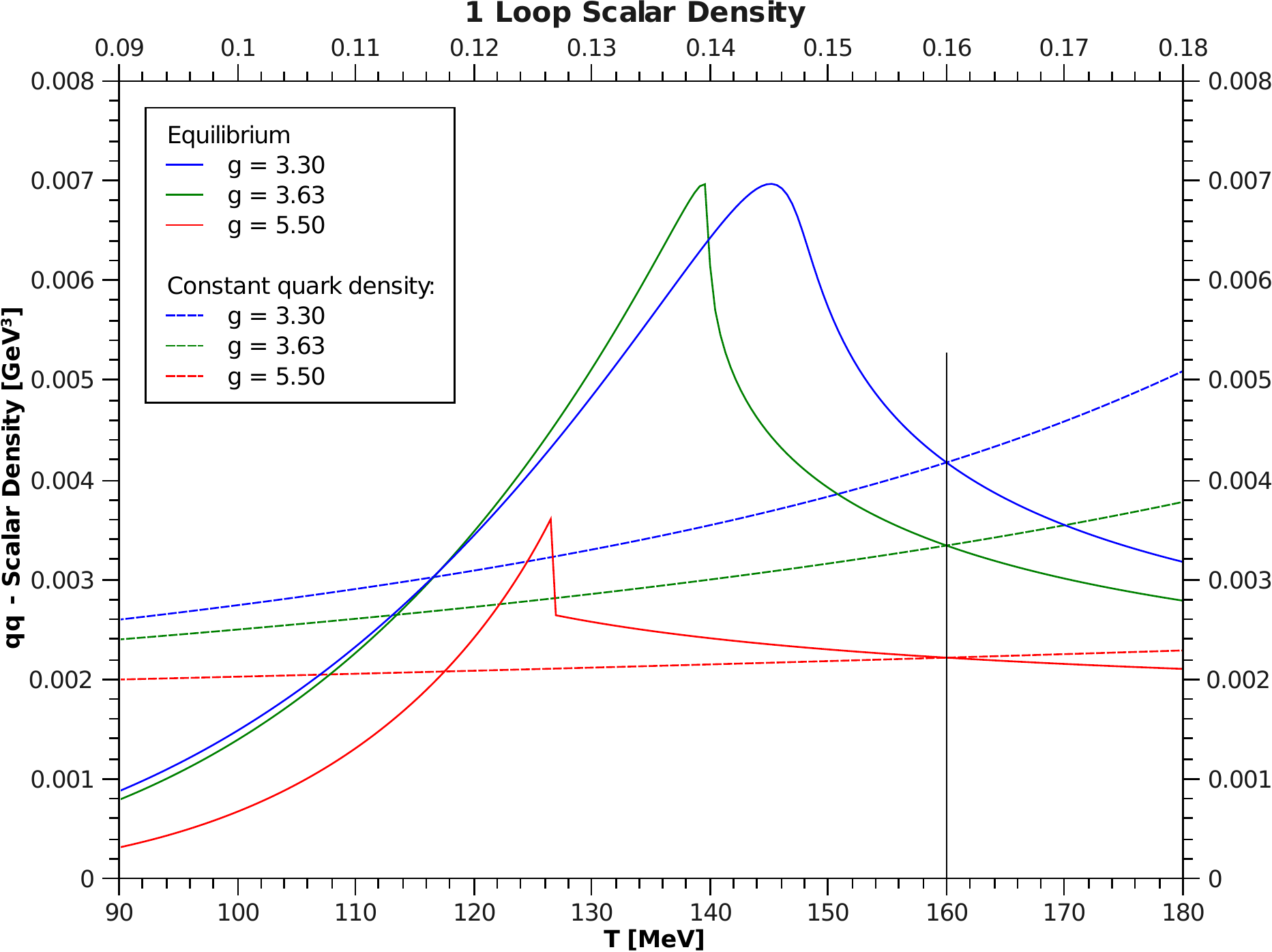}
  \caption{Comparison of the scalar density, $\langle \overline{\psi}
    \psi \rangle$, in case of thermal and chemical equilibrium (solid
    line) with the case of an expanding medium without chemical
    processes. Both simulated systems are in a box and cooled down by
    interactions with the heat bath. Without the chemical processes, no
    phase transition occurs, and the system is trapped in the
    chiral-symmetry restored phase.}
 \label{scalarDensity}
\end{figure}

Two interesting scenarios are shown in figure \ref{scalarDensity}. In
both scenarios, an initially equilibrated system is simulated in a
box. By interactions of the quarks with a thermal heat bath, the
system's temperature can be adjusted. The system in full chemical and
thermal equilibrium shows a phase transition at a given temperature. Due
to the lack of chemical processes, the second system does not show a
phase-transition behaviour any more. In both cases the system's thermal
energy changes but without a change in the quark density, the system is
trapped in its initial chiral state.

Extending the current model with chemical processes is not trivial if
total energy should be conserved. The Yukawa interaction allows for a
$\overline{\psi} \psi \to \sigma$ reaction. However, this process is not
a continuous one but acts simultaneously on the field and the
particles. While creating and annihilating the particles is a simple
numerical exercise, the quasi instantaneous particle-like interaction
with the filed is completely undefined. This problem is discussed in the
following section.

\section{Particle-wave interaction}
\label{sectionParticleWaveInteraction}

In the linear sigma model, particles and chiral fields are coupled via
the Yukawa like coupling. In our model, this is implemented via the
interaction potential in the Vlasov equation (\ref{5}). The gradient of
the field energy acts as a continuous force on the quarks, describing
the interaction between the quarks and the mean fields. This interaction
changes the quarks' momenta but lacks dissipation and entropy production
and thus can not drive the system to thermal equilibrium.

By including binary collisions (see section \ref{QuarkInteraction})
between the quarks, a kinetic thermalisation of the quarks can be
achieved, but so far the model lacks thermal and chemical equilibration
among the quarks and meson-mean fields, i.e., collective coherent
oscillations of the fields are not damped by the quark bath. In other
models \cite{nahrgang2011nonequilibrium, Herold:2013uza, Herold:2013bi},
the equations of motion are extended with a Langevin-like coupling. For
example, the $\sigma$ field is damped with a dissipative term, and
additionally a noise term is added. This allows for an effective
thermalisation of the meson-mean fields, but the quarks are treated as a
large (thermalised) reservoir, and the energy and momentum transfer from
the mean fields to the medium is neglected. The friction term in the
Langevin-mean-field equation thus leads to a continuous rescaling of the
quark energy, violating energy (and momentum) conservation.

To solve this problem we follow an ansatz motivated from kinetic
theory. Thus we treat the dissipation and noise of the mean fields as an
interaction between the quarks and the mean fields in a ``collision
like'' way, treating the interactions as local in time and with a
non-continuous energy-momentum transfer.

\subsection{Energy transfer between fields and particles}

The main problem of this approach is to find an effective way to
simulate such collision-like interactions between particles and mean
fields since the here used classical transport approach lacks a clear
interpretation in a kind of ``field-particle duality''. Here we use
energy and momentum as a common feature of particles and fields and
describe their exchange between particles and mean fields in a pragmatic
stochastical way.

The expressions for the energy of the mean fields follow from the
Lagrangian making use of Noether's theorem applied to the symmetry under
space-time translations, leading to the total energy and momentum of the
field in a spatial volume $V$,
\begin{alignat}{2}
\label{GKEnergy}
E &= \int_V \dd^3 \bvec{x} \ \epsilon(\bvec x) =\int_V \dd^3 \bvec{x} \left
  [ \frac{1}{2} \dot \phi^2 + \frac{1}{2} (\vec \nabla \phi)^2 + U(\phi)
\right ]\\
\label{GKMomentum}
\bvec{P} &= \int_V \dd^3 \bvec{x} \ \bvec{\Pi}(\bvec{x}) = \int_V \dd^3
\bvec{x} \dot{\phi}  \vec \nabla \phi.
\end{alignat}
To simulate a chemical annihilation of a quark pair,
\begin{equation}
  \psi + \bar \psi \to \hat \phi \to \phi
\end{equation}
we treat the annihilation as a $2 \to 1$ reaction of the $\psi \bar
\psi$ pair to an intermediate particle $\hat \phi$, which is defined via
its four-momentum. We map this four-momentum to an energy and momentum
transfer $\Delta E$ and $\Delta \bvec P$ to the meson-mean field,
$\phi(\bvec x, t)$.

For a given field $\phi(t)$, the equations of motion describe the
propagation of the field to a later time $\phi(t) \to \phi(t+\delta t)$.
Besides this normal propagation, we allow the field to suffer small
kicks $\delta \phi(x_k, t_k)$ at given discrete times $t_k$ and
positions $x_k$. These kicks can both increase or decrease the field's
energy, given by a variable $\Delta E(t)$.

The change of the field $\delta \phi$ is constrained by the
energy-momentum equations (\ref{GKEnergy}) and (\ref{GKMomentum}):
\begin{alignat}{2} 
\label{EDiff1}
 \Delta E(t_k) &= E \left [\phi(\bvec x, t_k) + \delta \phi(\bvec x, t_k)
 \right] - E \left [\phi(\bvec x, t_k) \right ], \\
\label{PDiff1}
\Delta \bvec P(t_k) &= \bvec P \left ( \phi(\bvec x, t_k) + \delta
  \phi(\bvec x, t_k) \right) - \bvec P \left (\phi(\bvec x, t_k) \right
).
\end{alignat}
In the general case, we can not find an inverse of this energy-momentum
functional, and we have to solve equations \ref{EDiff1} and
\ref{PDiff1}. In the (1+0)-di\-men\-sio\-nal case, a closed solution can
be found, if the potential $V(\bvec x)$ can be inverted.  In the
multi-dimensional case, there is no unique solution to
(\ref{EDiff1}-\ref{PDiff1}) anymore, and the field increment $\delta
\phi$ has to be parameterised. To obtain mathematical stability in the
equations of motion, the parameterisation has to be continuous,
non-pointlike and must have the same parametrical degrees of freedom as
equations (\ref{EDiff1}) and (\ref{PDiff1}) (two for the one-dimensional
case and four for the three-dimensional case). We have chosen a
travelling Gaussian wave packet as a parameterisation.

Even though it is numerically more challenging, we have two advantages
compared to the above described Langevin approaches: a discrete control
of the field energy change $\Delta E(t_k)$ allows for the microcanonical
description of a field-particle coupling without the need of an
uncontrolled heat bath. Secondly instead of a continuous and
deterministic $\dot \phi$ term, we can model a discrete and Monte-Carlo
like energy transfer $\Delta E(t)$.

To test our method, we were inspired by \cite{feynman1963theory} and
have investigated a simple but well understood model, the classical
damped harmonic oscillator coupled to a Langevin equation
\begin{equation}
  \frac{\partial^2}{\partial t^2} \phi(t) + \gamma
  \frac{\partial}{\partial t} \phi(t) + \phi(t) = \kappa \xi(t)
\end{equation}
with the usual dissipation-fluctuation relation for the noise term,
\begin{equation}
 \kappa = \sqrt{\frac{2 \gamma T}{\dd t}}
\end{equation}
These equations of motion describe a damped harmonic oscillator, which
is driven by a random force, where $\xi(t)$ is a Gaussian random
variable (``white noise'') which can increase or decrease the energy of
the system by 'kicking' the oscillator. On average, the oscillator will
show a Gaussian position distribution and an exponential energy
distribution $\propto \exp(-E/T)$.

To simulate the oscillator with our method, we need to model a $\Delta
E(t)$ for the processes $\sim \dot \phi(t)$ and $\kappa \xi(t)$. The
random kicks can be calculated by using the relation
\begin{equation}
 \frac{\dd E}{\dd t} = \dot \phi(t) F(t)
\end{equation}
which becomes
\begin{equation}
 \Delta E = \dot \phi(t) F(t) \Delta t
\end{equation}
for our numerical simulation, with $F(t)$ as the external force.

\begin{figure}[]
  \centering
  \includegraphics[width=0.7\textwidth]{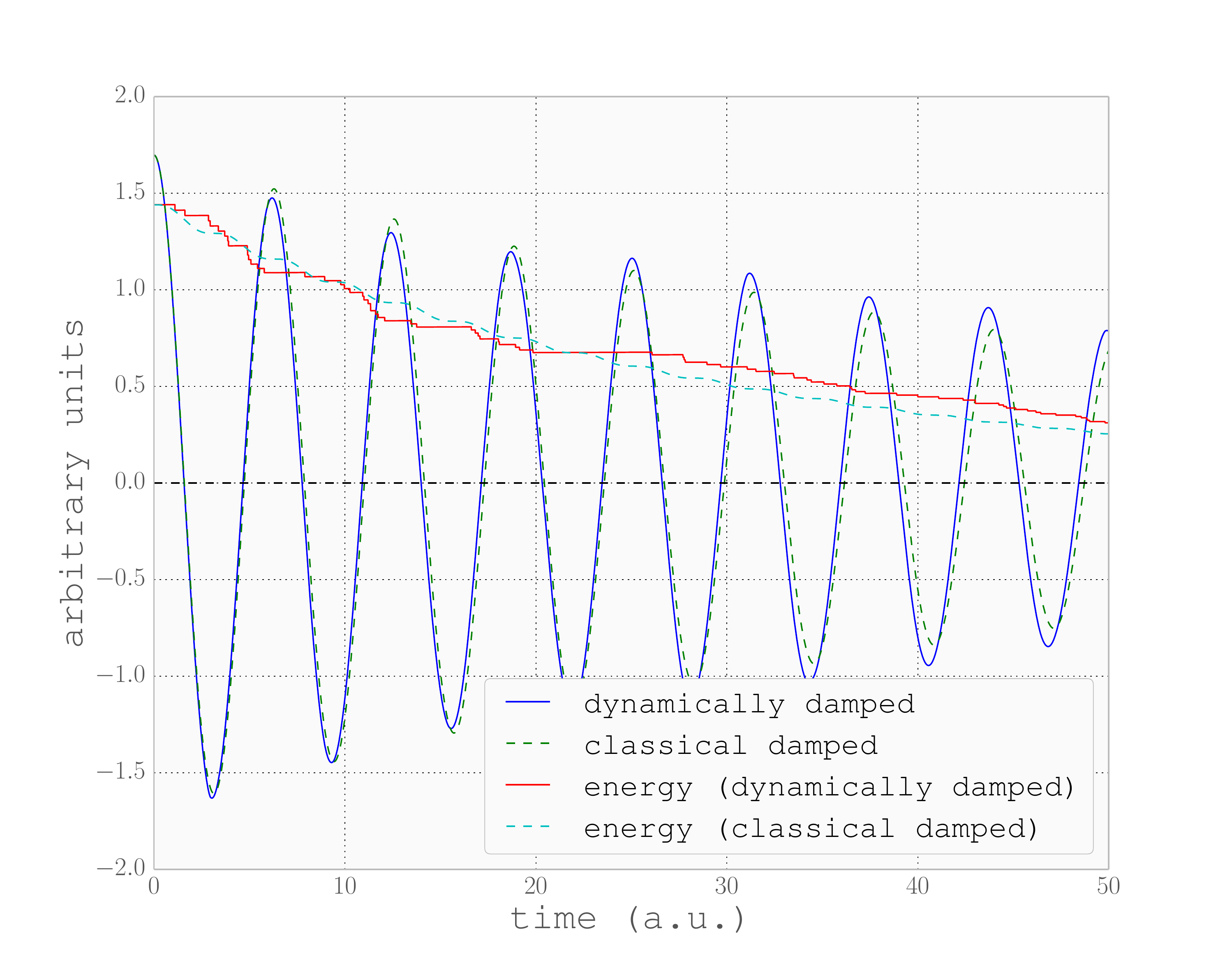}
  \caption{Comparison of two harmonic oscillators. The first one is
    classically damped, the second one has a statistical damping as
    described in this proceeding. In both cases, the average damping
    factor is $\gamma = 0.35$. The statistical method has a scaling of
    $N=50$. In the ensemble average, both show same decay-properties.}
 \label{fig1}
\end{figure}

Simulating the damping is more complicated, as we want to allow for
non-continuous damping. Therefore we come back to the simple damped
oscillator:
\begin{equation}
  \frac{\partial^2}{\partial t^2} \phi(t) + \gamma
  \frac{\partial}{\partial t} \phi(t) + \phi(t) = 0 
\end{equation}
The oscillator loses energy via the dissipative term $\gamma \dot
\phi$. In comparison, we simulate a harmonic oscillator with a dynamic
damping, modelled by a stochastic and discrete energy loss,
\begin{equation}
 \frac{\partial^2}{\partial t^2} \phi(t) + \phi(t) = \delta \phi(t),
\end{equation}
where $\delta \phi(t)$ is a little kick which changes the energy of the
system and is directly related to the external force $F(t)$. To simulate
discrete random kicks at times $t_k$ ($k \in \{1,2,\ldots,N\}$), leading
to energy changes $\Delta E_k$, we set
\begin{equation}
F(t)=\sum_{k=1}^{N} \frac{\Delta E_k}{\dot{\phi}(t)} \delta(t-t_k).
\end{equation}
The probability for the oscillator to lose a given amount of energy
$\Delta E$ is given by:
\begin{equation}
 P \left(\Delta E \right) = \gamma \cdot \dd t \cdot N/E_0
\end{equation}
The parameter $N$ can be seen as a `steppiness' scale of the system;
$E_0$ is the initial energy of the system, and the system can lose an
amount
\begin{equation}
 \Delta E = \frac{E_0}{N}
\end{equation}
in an interaction, leading to the same average energy loss as in the
classical case, which scales as
\begin{equation}
  \frac{\dd E}{\dd t} \sim e^{-2 \gamma t}.
\end{equation}
For $N \to \infty$, the original damped oscillator is restored, for
$N>1$, the oscillator's energy is damped via a discrete, statistically
decaying exponential function. This can be nicely seen in figure
\ref{fig1}.

Combining both the method of statistical energy gain and non-con\-ti\-nuous
damping, we can successfully describe a damped harmonic oscillator,
coupled to a Langevin force. Instead of a continuous and fully random
description, we can now control the energy flow in a discrete way. This
allows us to couple the oscillator or any other wave equation to a
finite, microcanonical heat bath or any other discrete system.

\section{Conclusion and outlook}

We have presented a dynamical, non-equilibrium simulation of a linear
$\sigma$-model with constituent quarks. Quarks, treated as test
particles, and the chiral fields, treated on the mean-field level, are
coupled via the corresponding mean-field equation. For additional
dynamics, a binary-collision term has been implemented for the
quarks. Elastic interactions allow an efficient thermalisation of the
quarks. However, this does not hold for the chiral fields. Coherent
oscillations are not damped, and the lack of chemical processes, both in
the quarks and the fields, does not allow for a consistent description
of a dynamical phase transition.

To treat this problem, we developed a method to allow a kinetic and
discrete Monte-Carlo like interaction between particles and classical
fields. As a proof of concept we have demonstrated that a consistent
mapping between the classical and continuous equations of motion for a
harmonic oscillator coupled to a Langevin force can be established. 

As the next step, we will apply this simulation method to the full
(3+1)-dimensional linear $\sigma$-model, which will finally enable us to
treat chemical reactions between the quarks and the chiral fields
consistently within our Vlasov-Boltzmann dynamics.

\textit{Acknowledgement.} This work has been supported by the German
Federal Ministry of Education and Research (BMBF F{\"o}rderkennzeichen \linebreak
05P12RFFTS). C.\ W.\ acknowledges support via the Helmholtz Research
School for Quark Matter Studies (H-QM). The authors are grateful to the
Center for Scientific Computing (CSC) at Frankfurt for the computing
resources.

%\bibliography{literature}

\begin{thebibliography}{1}
\providecommand{\url}[1]{\texttt{#1}}
\providecommand{\urlprefix}{URL }
\providecommand{\eprint}[2][]{\url{#2}}

\bibitem{gell1961nuovo}
M.~Gell-Mann and M.~Levy, \emph{Phys. Rev.} \textbf{122}, 345 (1961).

\bibitem{scavenius2001chiral}
O.~Scavenius, A.~Mocsy, I.~Mishustin, and D.~Rischke, \emph{Phys. Rev. C}
  \textbf{64}, 045202 (2001).

\bibitem{Xu:2004mz}
Z.~Xu and C.~Greiner, \emph{Phys. Rev. C} \textbf{71}, 064901 (2005),
  \eprint{hep-ph/0406278}.

\bibitem{nahrgang2011nonequilibrium}
M.~Nahrgang, S.~Leupold, C.~Herold, and M.~Bleicher, \emph{Phys. Rev. C}
  \textbf{84}, 024912 (2011).

\bibitem{Herold:2013uza}
C.~Herold, M.~Nahrgang, I.~Mishustin, and M.~Bleicher, \emph{PoS}
  \textbf{CPOD2013}, 021 (2013).

\bibitem{Herold:2013bi}
C.~Herold, M.~Nahrgang, I.~Mishustin, and M.~Bleicher, \emph{Phys. Rev. C}
  \textbf{87}, 014907 (2013), \eprint{1301.1214}.

\bibitem{feynman1963theory}
R.~P. Feynman and F.~L. Vernon, \emph{Ann. Phys.} \textbf{24}, 118 (1963).

\end{thebibliography}
%\bibliographystyle{appol-hvh}

\end{document}